\documentclass[%
prb,
amsmath,amssymb,
aps,
twocolumn,
]{revtex4-2}

\usepackage{graphicx}
\usepackage{dcolumn}
\usepackage{bm}
\usepackage{hyperref}

\newcommand\mk{\boldsymbol{\mathrm{k}}}
\newcommand{\bDelta}{\boldsymbol{\Delta}}
\newcommand{\rvec}[1]{\left| #1 \right\rangle}
\newcommand{\lvec}[1]{\left\langle #1 \right|}
\newcommand{\vecDot}[2]{\left \langle #1 \middle| #2 \right \rangle}
\newcommand{\abs}[1]{\left| #1 \right|}
\newcommand{\ti}[1]{\scriptscriptstyle{#1}}

\usepackage{ulem}
\usepackage{color}

\begin{document}

\preprint{APS/123-QED}

\title{Quantum geometric superfluid weight in multiband superconductors: A microscopic interpretation}

\author{Yi-Jian Hu}
	\address{Shenzhen Institute for Quantum Science and Engineering, Southern University of Science and Technology, Shenzhen 518055, Guangdong, China}
	\address{International Quantum Academy, Shenzhen 518048, China}
	\address{Guangdong Provincial Key Laboratory of Quantum Science and Engineering, Southern University of Science and Technology, Shenzhen 518055, China}
\author{Wen Huang}
	\email{huangw3@sustech.edu.cn}
	\address{Shenzhen Institute for Quantum Science and Engineering, Southern University of Science and Technology, Shenzhen 518055, Guangdong, China}
	\address{International Quantum Academy, Shenzhen 518048, China}
	\address{Guangdong Provincial Key Laboratory of Quantum Science and Engineering, Southern University of Science and Technology, Shenzhen 518055, China}


\date{\today}

\begin{abstract}
Even in non-interacting limit, electrons on different Bloch bands of a multiband system do not move as if they are oblivious to the presence of one another. Instead, they move in concert by virtue of a non-Abelian interband Berry connection. While the impact of this quantum geometric attribute manifests most famously through the Hall response of topological bands, the geometric effects in superconductors have attracted significant recent attention. In particular, much has been discussed about the quantum-metric-induced superfluid weight (SW) in flatband superconductors. In this study, we revisit the geometric SW in generic multiband superconductors and trace its origin to a series of microscopic processes. We separately derive the SW of models containing only intraband Cooper pairing and those involving interband pairing. Two classes of processes enabled by the so-called interband velocity (or the closely related interband Berry connection) are identified: one resembles the transfer of Cooper pairs between different bands, and the other corresponds to virtual single-electron back-and-forth tunneling between the bands. The former contribution manifests as effective Josephson coupling between the multiband superconducting order parameters, while the latter constitutes the only source of SW for a superconducting flatband well isolated from other bands. We further numerically evaluate the SW of a simple two-band superconductor with trivial band topology, showcasing how the geometric contribution is sensitive to the details of the multiband pairing configuration. In particular, we highlight an intriguing scenario of negative SW, which may pave way for the formation of novel pair density wave order. Our study provides deeper and more intuitive insight into the origin and nature of the SW induced by the quantum geometry of paired Bloch states. 
\end{abstract}

\maketitle


\section{\label{sec:intro}Introduction}
In semiclassical theories, the charge transport of Bloch electrons in solids are determined by their velocities. In a multiband system, the velocity operator, when expressed in the band basis in which the kinetic Hamiltonian is diagonalized, contains not only the usual group velocity (diagonal) terms, but also off-diagonal terms, denoted interband velocity. More concretely, the elements of the velocity matrix read~\cite{blount1962formalisms},
\begin{equation}\label{eq:velocity}
    V^{ij}_{\mu}=\delta_{ij}\partial_\mu\epsilon_i+ (\varepsilon_i-\varepsilon_j)\vecDot{\partial_{\mu}i}{j}\,.
\end{equation}
 Here, $\partial_\mu \equiv \partial_{k_\mu}$, and $\varepsilon_i$ and $\rvec{i}$ designate the energy dispersion of $i$-th Bloch band and the cell-periodic part of its wavefunction, respectively. Note that the implicit dependence on wavevector $\mk$ in each quantity is suppressed for notational brevity --- a practice we shall implement frequently throughout the paper. The presence of finite interband velocity signifies coherent transport, rather than totally decoupled motion, of electrons from different Bloch bands. Notice that $\mathrm{i}\vecDot{\partial_{\mu}i}{j}$ is the non-Abelian Berry connection between the two Bloch states, which encodes their quantum geometric characters. While the Berry connection (or the related interband velocity) by itself is not gauge invariant, geometric effects typically manifest in the form of their gauge-invariant products, such as $V^{ij}_\mu V^{ji}_\nu$, in the response functions. 
 
 The geometry of the individual Bloch states are formally described by the quantum geometric tensor~\cite{provost1980riemannian}, {\it e.g.},
 \begin{equation}
     \mathcal{R}_{\mu\nu}^i=\lvec{\partial_{\mu}i}\left(1-\rvec{i}\lvec{i}\right)\rvec{\partial_{\nu}i},
 \end{equation}
 whose imaginary and real parts give, respectively, the Berry curvature and quantum metric of band-$i$. The impact of the Berry curvature has been well studied, most notably in connection to the linear Hall response in the context of topological band theories~\cite{thouless1982quantized,niuqianRMp,qi-zhangRMP,hasan-kaneRMP}. For example, the Hall conductance of a classic two-band Chern insulator is unrelated to the band group velocities, but is expressed in terms of the interband velocity as~\cite{niu1985quantized}
\begin{eqnarray}
\sigma_{xy} &=& -i\frac{e^2}{h}\sum_{\mk} \frac{V^{12}_xV^{21}_y-V^{12}_yV^{21}_x}{(\varepsilon_{1}-\varepsilon_2)^2} \nonumber \\
&=& -i\frac{e^2}{h}\sum_{\mk} (\langle \partial_x 1|\partial_y 1\rangle - \langle \partial_{y} 1|\partial_x 1 \rangle) \,,
\end{eqnarray}
where the final expression contains an integral of the Berry curvature over momentum space. On the other hand, the effect of the quantum metric was much less discussed and has only recently begun to garner significant attention~\cite{gao2019nonreciprocal,mitscherling2020longitudinal,ahnRiemannianGeometryResonant2022,mitscherlingBoundResistivityFlatband2022,gao2023quantum,das2023intrinsic,huhtinen2023conductivity,torma2023essay,liu2024quantum}. \par

The geometric effects persist into the superconducting state of multiband superconductors. One novel example came to light recently is the superfluid weight (SW) of flatband superconductors~\cite{peottaSuperfluidityTopologicallyNontrivial2015,julkuGeometricOriginSuperfluidity2016,liangBandGeometryBerry2017,tovmasyan2016effective,huGeometricConventionalContribution2019,iskin2019superfluid,iskin2019origin,julkuSuperfluidWeightBerezinskiiKosterlitzThouless2020,xieTopologyBoundedSuperfluidWeight2020,wang2020quantum,herzog-arbeitmanManyBodySuperconductivityTopological2022,tormaSuperconductivitySuperfluidityQuantum2022,huhtinenRevisitingFlatBand2022,kitamura2022superconductivity,mao2023diamagnetic,mao2024upperbounds,iskin2024cooperpairing,jiang2024geometric,jiang2024superfluid}. Since electrons in flatbands have vanishing group velocity and are thus immobile, conventional theory has it that the SW must vanish when such systems develop Cooper pairing. If so, it would seem impossible to realize a flatband superconductor capable of maintaining superconducting phase coherence and supercurrent flow. However, interband velocity comes to the rescue by enabling effective transport of flatband electrons and therefore induce finite SW. More recently, it was also shown that, quantum geometry can facilitate the formation of novel phases of matter~\cite{hofmannSuperconductivityChargeDensity2023,kitamura2023quantum,kitamura2024spin}, such as pair density waves (PDW)~\cite{chenPairDensityWave2023,wang2024quantum,jiangPairDensityWaves2023,kitamuraQuantumGeometricEffect2022,ticea2024pair,sun2024flatbandfuldeferrelllarkinovchinnikovstatequantum,han2024quantum}, in strong-coupling multiband superconductors. Meanwhile, the geometric effects on the electromagnetic responses~\cite{chen2021quantum,ahn2021superconductivity,topp2021light,vermaOpticalSpectralWeight2021,ZhangJL2024,hu2024bandgeometricoriginsuperconductingdiode,tanaka2024nonlinear} and other properties~\cite{iskin2023extracting,yu2024non,chen2024Ginzburg-Landau,hu2024anomalouscoherencelengthsuperconductors,xiao2024quantumgeometriceffectshiggs} of more general multiband superconductors have also received much interest.\par
\begin{figure*}
    \includegraphics[scale=0.60]{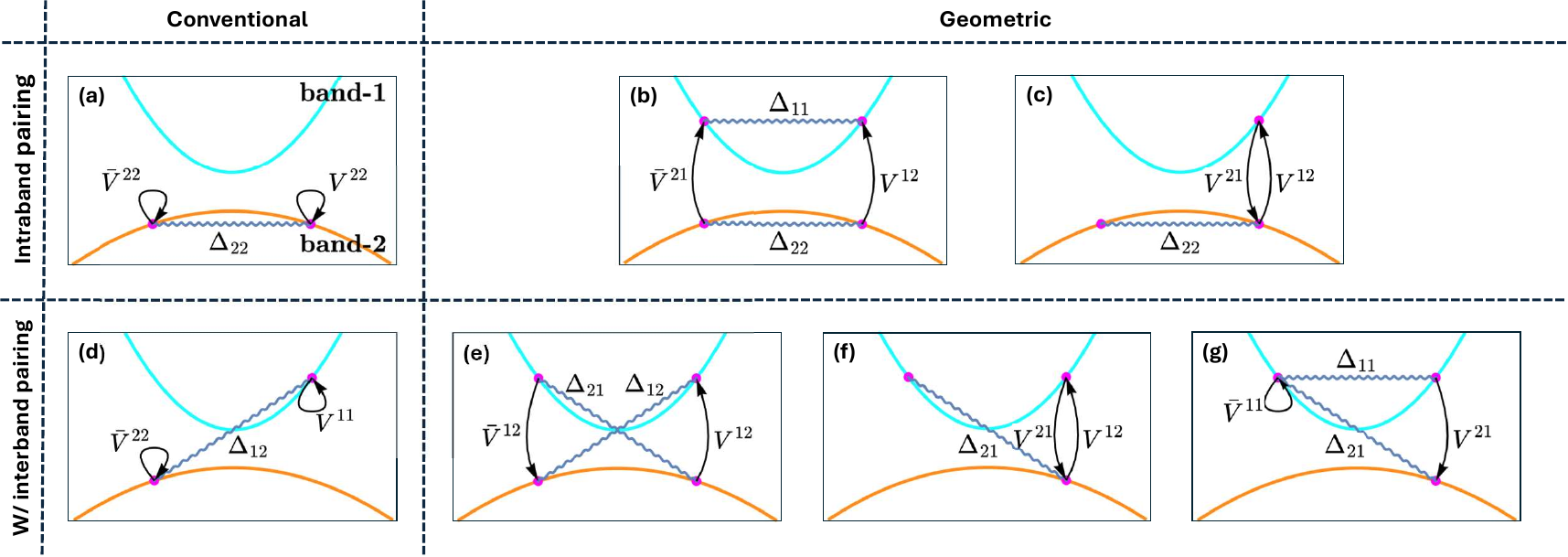}
    \caption{Illustration of the microscopic processes that generate superfluid weight in multiband superconductors. The cyan and orange curves sketch the dispersion of a two-band model constructed from a two-orbital model with nonvanishing orbital mixing. Cooper pairing is represented by wavy lines connecting electron pairs. The curved arrows indicate the tunneling of electrons enabled by intraband or interband velocities (or interband Berry connection). The element of the velocity operator $V^{ij}$ ($V^{ij}_\mu$) is identified with $V^{ij}_{\mu\mk}c^\dagger_{i\mk}c_{j\mk}$ and is interpreted as the tunneling of an electron at wavevector $\mk$ from band-$j$ to band-$i$. In correspondence, $\bar{V}^{ij}$ ($\bar{V}^{ij}_\mu$) is identified with $-(V^{ij}_{\mu\bar{\mk}})^\ast c_{i\bar{\mk}}c^\dagger_{j\bar{\mk}}$ and depicts the transfer of a hole at wavevector $-\mk$ from band-$j$ to band-$i$, or equivalently, the transfer of an electron at this wavevector from band-$i$ to band-$j$. (a-c) involve only intraband Cooper pairing, while (d-g) also take into account interband pairing. (a) and (d) illustrate conventional Cooper pair tunneling processes described by Eq.~\eqref{eq:conv0} and Eq.~\eqref{eq:processd}, respectively. (b),~(e) and (g) depict the Cooper pair transfer associated respectively with Eq.~\eqref{eq:geo0}, Eq.~\eqref{eq:processe} and Eq.~\eqref{eq:processg}. These processes induce effective Josephson coupling between the two pairing order parameters involved. For example, (b) amounts to a coupling $\Delta_{11}^\ast\Delta_{22}$. (c) and (f) describe the back-and-forth interband single-electron tunneling originating from Eq.~\eqref{eq:processc} and Eq.~\eqref{eq:processf}, respectively. Notably, if one of the bands is a flatband and if it is the only band developing Cooper pairing, then (c) constitutes the only contribution to the superfluid weight. Also note that conventional contributions described by Eqs.~\eqref{eq:deltaconv} and \eqref{eq:partialD12} are not sketched in this figure.}
    \label{fig:sketch}
\end{figure*}
While the quantum-geometry-induced SW has been studied in great depth, most studies have concentrated on flatband systems. Moreover, in most cases the geometric effect is interpreted in terms of the somewhat abstract mathematical object of quantum geometric tensor. Little is known about the underlying physical processes that generate the geometric SW. In this paper, we revisit the problem and provide an intuitive microscopic interpretation on the basis of the interband velocity or, the closely related interband Berry connection. In this description, the conventional SW measures the ability of Cooper pairs to `transfer' by virtue of the intraband velocity, {\it i.e.}, the diagonal elements of Eq.~\eqref{eq:velocity}. In similar spirit, the geometric SW can be traced to two classes of processes enabled by interband velocity (or interband Berry connection): one resembles Cooper pair transfer between different bands, and the other is single-electron tunneling from one band to another and then back. Figure~\ref{fig:sketch} summarizes most of the microscopic processes, both conventional and geometric, we identify in this work. For completeness, we also include in our derivation interband Cooper pairing. Therefore, our analyses are expected to capture some qualitative features of models with strong Cooper pairing.\par 
Our study goes beyond the analyses of Ref.~\onlinecite{kitamuraQuantumGeometricEffect2022}, with which our analytical results are in overall agreement, except for some subtle but important differences. The most significant further step we achieve here is to elucidate the underlying microscopic processes that generate the SW and to interpret the geometric SW in a more tangible language. \par
The rest of this paper is organized as follows. In Sec.~\ref{sec:formalism}, we present the general formalism for deriving the SW within linear response theory. In Sec.~\ref{sec:intra}, we derive the SW for models without interband pairing, elucidate the physical origin and classify
the microscopic process associated with each of the contributing terms. Sec.~\ref{sec:inter} includes the interband pairing in the derivation, outlining several new contributions. In Sec.~\ref{sec:model}, we perform numerical calculations of the SW of a representative two-band model, demonstrating strong sensitivity the geometric contribution has on the multiband pairing configuration. We also highlight an exotic scenario with negative total SW, which may provide novel intrinsic mechanism for the formation of PDW order.  Sec.~\ref{sec:summary} summarizes this study.




\section{Formalism}\label{sec:formalism}
For simplicity, let us consider a spin-singlet superconducting state in a multiband model without spin-orbit coupling, and assume that time-reversal symmetry is preserved. Typically, one starts with a tight-binding model containing multiple orbital (or sublattice) degrees of freedom, wherein the kinetic and the pairing terms of the Bogoliubov-de Gennes (BdG) Hamiltonian are both written in the orbital Nambu-spinor basis $(c_{A\mk\uparrow},c_{B\mk\uparrow},...,c^\dagger_{A\bar{\mk}\downarrow},c^\dagger_{B\mk\downarrow},...)^T$,
\begin{equation}
    \mathcal{H}_{\mk}^{\mathbf{BdG}}=
    \begin{pmatrix}
        \tilde{\mathcal{H}}_{\mk\uparrow} & \tilde{\bDelta} \\
        \tilde{\bDelta}^{\dagger} & -\tilde{\mathcal{H}}_{\bar{\mk}\downarrow}^\ast
    \end{pmatrix} \,,
    \label{eq:Hbdg}
\end{equation}
where $A,B,...$ label orbital indices, $\bar{\mk}=-\mk$, $\tilde{\mathcal{H}}_{\mk \uparrow} = \tilde{\mathcal{H}}_{\bar{\mk} \downarrow}^*=\tilde{\mathcal{H}}_{\mk}$ is the normal-state Hamiltonian and $\tilde{\bDelta}$ (with implicit $\mk$-dependence) represents the pairing matrix. Here and after, we use the tilde symbol to distinguish quantities in orbital-basis representation from those in the band basis to be presented later. 

While the SW can be derived from the gradient expansion of the grand potential, here we employ the standard linear response formalism according to Ref.~\onlinecite{liangBandGeometryBerry2017}, wherein the dependence on interband velocity is evident from the outset. The SW is given by~\cite{liangBandGeometryBerry2017,chenPairDensityWave2023}:
\begin{equation}\label{eq:Doriginal}
\begin{aligned}
D_{\mu \nu}=T_{\mu \nu}+\Pi_{\mu \nu}=\frac{1}{\beta} \sum_{\omega_n, \mk} \operatorname{Tr}\left\{\tilde{m}_{\mu \nu}^{-1} \mathcal{G}+\mathcal{G}\tilde{V}_{\mu} \mathcal{G}\tilde{V}_{\nu}\right\},
\end{aligned}
\end{equation}
where $\beta$ is the inverse of temperature $T$, $\omega_n =(2n+1)\pi/\beta$ denotes fermionic Matsubara frequency and $\mathcal{G} \equiv \mathcal{G}(\mathrm{i}\omega_n,\mk)=(i\omega_n - \mathcal{H}_{\mk}^\text{BdG})^{-1}$ is the Gor'kov Green's function associated with the BdG Hamiltonian. Here and below, the summation over $\mk$, $\sum_{\mk}$, is always understood as $\frac{1}{Z}\sum_{\mk}$, where $Z$ denotes the size of the system. 

The two terms in the above expression, $T_{\mu\nu}$ and $\Pi_{\mu\nu}$, correspond respectively to the diamagnetic and paramagnetic contributions. The inverse mass tensor $\tilde{m}_{\mu \nu}^{-1}$ and velocity operator $\tilde{V}_{\mu}$ are defined as follows,
\begin{align}
\tilde{m}_{\mu\nu}^{-1}&=\left.\frac{\partial^{2}\mathcal{H}_{\mk}^{\mathrm{BdG}}}{\partial_{\mu}\partial_{\nu}}\right|_{\Delta\to 0},\\
\tilde{V}_{\mu}&=\left.\tau_z \partial_{\mu}\mathcal{H}_{\mk}^{\mathrm{BdG}}\right|_{\Delta\to 0}.\label{eq:Vorbital}
\end{align}
Here $\tau_i $ with $i=x,y,z$ denotes the Pauli matrix acting in the Nambu particle-hole space.\par

Using integration by parts and utilizing the periodic boundary condition in the Brillouin zone, the diamagnetic term can be made into
\begin{equation}
T_{\mu\nu}=-\frac{1}{\beta} \sum_{\omega_n, \mk} \operatorname{Tr}\left[\mathcal{G}\partial_{\mu} \mathcal{H}_{\mk}^{\mathrm{BdG}} \mathcal{G}\left(\left.\partial_{\nu} \mathcal{H}_{\mk}^{\mathrm{BdG}}\right|_{\Delta \to 0}\right)\right],
\label{eq:diag1}
\end{equation}
where
\begin{equation}
\partial_{\mu} \mathcal{H}_{\mk}^{\mathrm{BdG}}=\left.\partial_{\mu} \mathcal{H}_{\mk}^{\mathrm{BdG}}\right|_{\Delta\to 0}+\begin{pmatrix}
    0 & \partial_{\mu}\tilde{\bDelta} \\
    \partial_{\mu}\tilde{\bDelta}^{\dagger} & 0
\end{pmatrix}.
\label{eq:dHbdg}
\end{equation}
For later convenience, we shall refer to the SW contribution involving the $\mk$-gradient of $\tilde{\bDelta}$ in Eq.~\eqref{eq:dHbdg} as the ``$\Delta$-gradient contribution", and call the remaining SW in $T_{\mu\nu}$ and $\Pi_{\mu\nu}$ the "kinetic-gradient contribution" in view of the fact that it is related to the $\mk$-gradient of the kinetic part of the Hamiltonian. They are given respectively by
\begin{equation}
\label{eq:Dkindelta}
    \begin{aligned}
        D_{\mu\nu}^{\text{kin}}&=\frac{1}{\beta} \sum_{\omega_n, \mk} \operatorname{Tr}\left[\mathcal{G}\tilde{V}_{\mu}\mathcal{G}\tilde{V}_{\nu}-\mathcal{G}\tau_z \tilde{V}_{\mu}\mathcal{G}\tau_z \tilde{V}_{\nu} \right], \\
        D_{\mu\nu}^{\Delta}&=-\frac{1}{\beta} \sum_{\omega_n, \mk} \operatorname{Tr}\left[\mathcal{G} \left(
    \begin{array}{cc}
       0  & \partial_{\mu}\tilde{\bDelta} \\
       \partial_{\mu}\tilde{\bDelta}^{\dagger}  & 0
    \end{array}
    \right) \mathcal{G} \tau_z \tilde{V}_{\nu} \right].
    \end{aligned}
\end{equation}
If the orbital-basis pairing matrix $\tilde{\bDelta}$ is $\mk$-independent, or if one works in the weak-coupling limit where the pairing gap is much smaller than any other kinetic and interaction energy scales, the $\Delta$-gradient contribution can be neglected.

Everything so far has been presented in the orbital-basis language. However, this representation has certain shortfall, as the pairing matrix thus-constructed does not in general reflect the true scheme of Cooper pairing in a multiband system. This is because, when projected into the band-basis representation in which the kinetic part of the Hamiltonian is diagonalized, such a construction generally encapsulates comparable intraband and interband Cooper pairings with the exception of some rare cases such as when $ \tilde{\bDelta}$ in Eq.~\eqref{eq:Hbdg} is an identity matrix. In reality, however, interband pairing is typically smaller in dispersive multiband systems due to the absence of an inherent interband Cooper instability. Hence, we shall turn to the band-basis description, which allows us to distinguish contributions associated with intraband and interband pairings. 

Let $\rvec{i}~(\rvec{\bar{i}})$ and $\varepsilon_{i}~(\bar{\varepsilon}_{i})$ be the eigenvector and corresponding eigenvalue of the orbital-basis particle (hole) Hamiltonian $\tilde{\mathcal{H}}_{\mk\uparrow}~(\tilde{\mathcal{H}}^{*}_{\bar{\mk}\downarrow})$, where $i=1,2,...,N$ label the band indices. Then the wavefunction of the Bogoliubov quasiparticles $\rvec{\psi_a}$ can be expanded as
\begin{equation}
    \rvec{\psi_a}=\sum_{i=1}^{N}\left( \phi_{ai}^+\rvec{+,i}+\phi_{ai}^{-}\rvec{-,\bar{i}} \right).
    \label{eq:psia}
\end{equation}
Here $a=1,2,...,2N$ are BdG band indices. $\rvec{\pm,i}$ are defined as $\rvec{\pm}\otimes\rvec{i}$ with $\rvec{+}=(1,0)^T$ and $\rvec{-}=(0,1)^T$ denoting particle and hole subspace, respectively. $\phi_{ai}^{\pm}$ are the expand coefficients. Furthermore, let $E_{a\mk}$ denote the BdG energy dispersion, then the Gor'kov Green's function can be written as
\begin{equation}
\mathcal{G}_{\mk}(\mathrm{i}\omega_n)=\frac{1}{\mathrm{i}\omega_n-\mathcal{H}^\text{BdG}_{\mk}}=\sum_{a=1}^{2N}
\frac{\rvec{\psi_a}\lvec{\psi_a}}{\mathrm{i} \omega_n-E_{a\mk}}.
\label{eq:Green}
\end{equation}
Substitute Eq.~\eqref{eq:psia} into it one can also obtain the Green's function in band basis.

The velocity operator Eq.~\eqref{eq:Vorbital} must also be transformed into band basis accordingly. This can be done separately for the particle and hole subspace, getting, in the Nambu-spinor basis $(c_{1\mk\uparrow},c_{2\mk\uparrow},...,c^\dagger_{1\bar{\mk}\downarrow},c^\dagger_{2\bar{\mk}\downarrow},...)^T$,
\begin{equation}\label{eq:Vband}
V_\mu^{b} = \begin{pmatrix}
V_\mu  &  0 \\
0  &  \bar{V}_\mu 
\end{pmatrix}
\end{equation}
and the matrix elements of the two parts are given as follows
\begin{eqnarray}\label{eq:Velocityij}  
V_{\mu}^{ij} &\equiv & \lvec{i}\partial_{\mu}\tilde{\mathcal{H}}_{\mk}\rvec{j} \nonumber \\
    &=& \delta_{ij}\partial_{\mu}\varepsilon_{i}+(\varepsilon_{i}-\varepsilon_{j})\vecDot{\partial_{\mu}i}{j}, \nonumber\\
    \bar{V}_{\mu}^{ij}&\equiv& \lvec{\bar{i}}\partial_{\mu}\tilde{\mathcal{H}}^{*}_{\bar{\mk} }\rvec{\bar{j}}\nonumber \\
    &=&\delta_{ij}\partial_{\mu}\bar{\varepsilon}_{i}+(\bar{\varepsilon}_{i}-\bar{\varepsilon}_{j})\vecDot{\partial_{\mu}\bar{i}}{\bar{j}}.
\end{eqnarray}
By time-reversal invariance, one has $\varepsilon_i=\bar{\varepsilon}_i$, $V^{ii}_{\mu}=\bar{V}^{ii}_{\mu}$ and $V^{ij}_{\mu}=\bar{V}^{ij}_{\mu}$. The velocity operator is also Hermitian, satisfying $V^{ij}_{\mu}=(V^{ji}_{\mu})^*$. Further, it is important to formally identify $V_{\mu}^{ij}$ with $V_{\mu\mk}^{ij}c^\dagger_{i\mk\uparrow}c_{j\mk\uparrow}$ and $\bar{V}_{\mu}^{ij}$ with $ -(V_{\mu\bar{\mk}}^{ij})^\ast c_{i\bar{\mk}\downarrow}c^\dagger_{j\bar{\mk}\downarrow}$, which can be interpreted as the transfer of a spin-up electron (spin-down hole) from band-$j$ to band-$i$. 

The pairing function must be transformed into band basis too. Let us first denote the pairing matrix elements in band-basis $\Delta_{ij}$. They are related to the orbital-basis pairing $\tilde{\bDelta}$ through $\Delta_{ij}=\lvec{i}\tilde{\bDelta}\rvec{\bar{j}}$. One hence obtains $\tilde{\bDelta}=\sum_{i,j}\Delta_{ij}\rvec{i}\lvec{\bar{j}}$, from which we have,
\begin{equation}
\partial_{\mu}\tilde{\bDelta}=\sum_{i,j}\left( \Delta_{ij}\rvec{\partial_{\mu}i}\lvec{\bar{j}}+
\Delta_{ij}\rvec{i}\lvec{\partial_{\mu}\bar{j}}+
\partial_{\mu}\Delta_{ij}\rvec{i}\lvec{\bar{j}}
\right).
\label{eq:dDo}
\end{equation}
Substituting the band-basis quantities into Eq.~\eqref{eq:Dkindelta}, we can rewrite the SW as
\begin{widetext}
\begin{align}
    &D_{\mu\nu}^{\text{kin}}=2\sum_{\mk}\sum_{ijlm}C_{ijlm}^{\ti{++--}}\left(V_{\mu}^{ij}\bar{V}_{\nu}^{lm}+V^{ij}_{\nu}\bar{V}^{lm}_{\mu}\right),
    \label{eq:sw_Dkin} \\
    &D_{\mu\nu}^{\Delta}=-2\text{Re}\sum_{\mk}\sum_{ijlm\sigma} C_{ijlm}^{\ti{+-\sigma\sigma}}\left( \delta^{+\sigma}V^{lm}_{\nu}-\delta^{-\sigma}\bar{V}^{lm}_{\nu} \right)  \left[\sum_{p}\left( \Delta_{ip}\vecDot{\partial_{\mu}\bar{p}}{\bar{j}}+\vecDot{i}{\partial_{\mu}p}\Delta_{pj} \right)+\partial_{\mu}\Delta_{ij}\right] 
     \label{eq:sw_Ddelta} \,,
\end{align}
\end{widetext}
where
\begin{equation}
    C_{ijlm}^{\ti{\sigma_1\sigma_2\sigma_3\sigma_4}}=\sum_{a,b}\frac{f(E_a)-f(E_b)}{E_a-E_b}\phi_{ai}^{\sigma_1*}\phi_{bj}^{\sigma_2}\phi_{bl}^{\sigma_3*}\phi_{am}^{\sigma_4} \,,
\end{equation}
$f(E)$ is the Fermi distribution function, $\sigma=\pm$, and $\delta^{+\sigma}(\delta^{-\sigma})=1$ if $\sigma=+(-)$, or else it equals zero. Notice that the coefficients satisfy the relation $(C_{ijlm}^{+-\sigma\sigma})^*=C_{jiml}^{-+\sigma\sigma}$. We note that similar expressions were also obtained in Ref.~\onlinecite{kitamuraQuantumGeometricEffect2022}. However, unlike that study, we hereby explicitly distinguish particle- and hole-like quantities, which shall allow for clear identification of the microscopic processes underlying each contribution. Moreover, the square bracket of Eq.~\eqref{eq:sw_Ddelta} encloses two interband Berry connection terms that correspond to geometric contribution, which were not recognized in Ref.~\onlinecite{kitamuraQuantumGeometricEffect2022}. 

Below, for the sake of a clearer presentation, we shall first derive the SW for models without interband pairing, and only turn to discuss SW related to interband pairing afterwards. \par
 

\section{Intraband pairing induced superfluid weight}\label{sec:intra}
This section is the central part of this study. Here, we provide a detailed derivation of the SW in multiband superconductors with only intraband Cooper pairings present. We shall showcase how microscopic processes enabled by the interband velocity (or interband Berry connection) induce SW, thereby providing an intuitive understanding of the quantum geometric effects on the nature of the superfluidity in these systems.

\subsection{Kinetic-gradient contribution}\label{sec:SWkin}
In the absence of interband pairing, it is straightforward to diagonalize the subblocks of the BdG Hamiltonian associated with the individual bands. The Bogoliubov quasiparticle energy spectrum for band-$i$ is given by $\pm E_{i}=\pm\sqrt{\epsilon_{i}^2 +\abs{\Delta_{ii}}^2}$, with corresponding wavefunctions $\rvec{\psi_i^{+}}=u_i\rvec{+,i}+v_i\rvec{-,\bar{i}}$, $\rvec{\psi_i^{-}}=-v_i^*\rvec{+,i}+u_i^*\rvec{-,\bar{i}}$. The coefficients $u_i$ and $v_i$, again dropping the implicit $\mk$-dependence, are given by
\begin{equation}
u_{i}=\frac{1}{\sqrt{2}} \sqrt{1+\frac{\epsilon_{i}}{E_{i}}},~~~v_{i}=\frac{1}{\sqrt{2}}\frac{\Delta_{ii}^*}{\abs{\Delta_{ii}}} \sqrt{1-\frac{\epsilon_{i}}{E_{i}}} .
\label{eq:uv}
\end{equation}
Substituting these into Eq.~\eqref{eq:sw_Dkin} we get the total kinetic-gradient contribution, which can be separated into conventional and geometric parts, 
\begin{equation}\label{eq:sw_indk}
D^\text{kin}_{\mu\nu}=D_{\mu\nu,conv}^{\text{kin}}+D_{\mu\nu,geo}^\text{kin}.
\end{equation}
The conventional term contains only intraband velocities,
\begin{equation}\label{eq:conv0}
\begin{aligned}
    D_{\mu\nu,conv}^\text{kin}&=2\sum_{\mk,i}\frac{|u_iv_i|^2}{E_i} (V_{\mu}^{ii}\bar{V}_{\nu}^{ii}+\bar{V}_{\mu}^{ii}V_{\nu}^{ii})\\
    &=\sum_{\mk,i}\frac{\abs{\Delta_{ii}}^2}{(E_{i})^3}V_{\mu}^{ii}V_{\nu}^{ii} \,. \\ 
\end{aligned}
\end{equation}
Here, we have used the relation $V^{ii}_{\mu}=\bar{V}^{ii}_{\mu}$. This term is positive definite. In the weak-coupling and continuum limit, the diagonal elements of $D_{\mu\nu,conv}^\text{kin}$ reduce to the well-known result of $\frac{\rho}{m^\ast}$, where $\rho$ and $m^\ast$ denote the carrier density and the effective mass, respectively. They vanish in the case of flat bands, for which $m^\ast=\infty$. 

There is an intuitive microscopic interpretation of the conventional SW Eq.~\eqref{eq:conv0}, as it can be regarded as originating from an effective intraband Josephson coupling. Here, the intraband velocities $V_{\mu}^{ii}$ and $\bar{V}_{\nu}^{ii}$ (identified as $V_{\mu\mk}^{ii}c^\dagger_{i\mk\uparrow}c_{i\mk\uparrow}$ and $-(V_{\nu\bar{\mk}}^{ii})^\ast c_{i\bar{\mk}\downarrow}c^\dagger_{i\bar{\mk}\downarrow}$, respectively) act to respectively tunnel a spin-up electron at $\mk$ and a spin-down hole at $-\mk$, both of band-$i$, to the same band. These processes are equivalent to the transfer of a Cooper pair of electrons at $\mk$ and $-\mk$ to the same band, as schematically drawn in Fig.~\ref{fig:sketch}(a). By analogy with the standard Josephson coupling which is induced by tunneling of Cooper pairs between different superconductors, the above process can be viewed as an effective intraband Josephson coupling. Such a microscopic point of view provides an illuminating perspective to understanding the origin of the geometric SW to be discussed below.


The second term of Eq.~\eqref{eq:sw_indk} is of pure quantum-geometric origin, as it contains only interband velocities and is absent in traditional theories. It reads,
\begin{equation}
\begin{aligned}
    D_{\mu\nu,geo}^\text{kin}&=4\sum_{\mk,i\neq j}\frac{u_i^* v_i u_j v_j^*}{E_i +E_j}\left( V^{ij}_{\mu}\bar{V}^{ji}_{\nu}+V^{ij}_{\nu}\bar{V}^{ji}_{\mu} \right)\\
    & = \sum_{\mk,i\neq j}\frac{\Delta_{ii}^*\Delta_{jj}}{E_iE_j(E_i +E_j)}\left( V^{ij}_{\mu}\bar{V}^{ji}_{\nu}+V^{ij}_{\nu}\bar{V}^{ji}_{\mu} \right).
\end{aligned}
\label{eq:geo0}
\end{equation}
Following the above description, $V_{\mu}^{ij}$ tunnels a spin-up electron at $\mk$ from band-$j$ to band-$i$, and $\bar{V}_{\nu}^{ji}$ tunnels a spin-down hole at $-\mk$ from band-$i$ to band-$j$ or, the inversed process with spin-down electron. Put together, as schematically depicted in Fig.~\ref{fig:sketch} (b), $V^{ij}_{\mu}\bar{V}^{ji}_{\nu}$ and $V^{ij}_{\nu}\bar{V}^{ji}_{\mu}$ can be viewed as the transfer of a Cooper pair of electrons from band-$j$ to band-$i$. These processes induce an effective interband Josephson coupling between the two intraband order parameters $\Delta_{ii}$ and $\Delta_{jj}$, which underlies the microscopic origin of this geometric SW.

We note that this geometric SW is finite only if band-$i$ and band-$j$ both develop pairing. And it is generally small in the weak-coupling limit where $\Delta/E_F \ll 1$, because the coherence factors $u_i^\ast v_i$ and $u_jv_j^\ast$ in general do not peak at the same wavevectors, as was also noted in Ref.~\cite{chenPairDensityWave2023}. However, it can become significant if the two superconducting bands lie in proximity to the Bose-Einstein-Condensation (BEC) limit or the flatband limit where $\Delta/E_F \gtrsim 1$. It could also acquire opposite signs, depending on the relative phase between $\Delta_{ii}$ and $\Delta_{jj}$. 

A remark about the relation of the above contribution to quantum geometry is in order. Assume real order parameters for simplicity, the geometric term Eq.~\eqref{eq:geo0} can be written as
\begin{equation}
    D_{\mu\nu,geo}^\text{kin}=\sum_{\mk,i\neq j}\frac{\Delta_{ii}\Delta_{jj}}{E_i E_j(E_i +E_j)}\left(V^{ij}_{\mu}V^{ji}_{\nu}+V^{ji}_{\mu}V^{ij}_{\nu}\right),
    \label{eq:geo0trs}
\end{equation}
where
\begin{equation}\label{eq:VV}
V^{ij}_{\mu}V^{ji}_{\nu}+V^{ji}_{\mu}V^{ij}_{\nu} = 2(\varepsilon_{i}-\varepsilon_{j})^2\text{Re}\left(\vecDot{\partial_{\mu}i}{j} \vecDot{j}{\partial_\nu{i}} \right) \,.
\end{equation}

In a two-band system, one has $|j\rangle\langle j| = 1-|i\rangle\langle i |$, ($i\neq j$), hence 
\begin{equation}\label{eq:QG1CooperTunnel}
\vecDot{\partial_{\mu}i}{j} \vecDot{j}{\partial_\nu{i}} = \vecDot{\partial_{\mu}i}{\partial_\nu{i}} - \vecDot{\partial_{\mu}i}{i} \vecDot{i}{\partial_\nu{i}} = \mathcal{R}^i_{\mu\nu} \,,
\end{equation}
which is the quantum geometric tensor of the Bloch band-$i$. Hence, the r.h.s.~of Eq.~\eqref{eq:VV} and therefore the geometric SW Eq.~\eqref{eq:geo0trs} are directly related to the real part of the geometric tensor -- the quantum metric~\cite{provost1980riemannian}, $\text{Re}\left(\mathcal{R}^i_{\mu\nu}\right)$. 

However, for systems consisting of more than two bands, Eq.~\eqref{eq:geo0trs} in general cannot be expressed in terms of the quantum metric of the individual Bloch bands. Nonetheless, the dependence on the geometric property of the Bloch bands is still explicit in the gauge-invariant products of interband velocities, $V^{ij}_{\mu}V^{ji}_{\nu}$ or, that of the non-Abelian Berry connection between the Bloch bands. 

\subsection{$\Delta$-gradient contribution}\label{sec:SWdelta}
Similar to the previous subsection, the $\Delta$-gradient contribution Eq.~\eqref{eq:sw_Ddelta} also consists of conventional and geometric parts,
\begin{equation}
    D^{\Delta}_{\mu\nu}=D^{\Delta}_{\mu\nu,conv}+D^{\Delta}_{\mu\nu,geo}.
\end{equation}
Utilizing ~Eq.~\eqref{eq:uv} and after some algebra, one obtains,
\begin{widetext}
    \begin{align}
    D^{\Delta}_{\mu\nu,conv}&=\text{Re}\sum_{i,\mk}\frac{\abs{u_i}^2-\abs{v_i}^2}{E_i}{}u_{i}^*v_i(V^{ii}_{\nu}+\bar{V}^{ii}_{\nu})(\Delta_{ii}\vecDot{\partial_{\mu}i}{i}-\Delta_{ii}\vecDot{\partial_{\mu}\bar{i}}{\bar{i}}-\partial_{\mu}\Delta_{ii})
    = -\text{Re}\sum_{i,\mk}\frac{\varepsilon_i}{E_i^3} \Delta_{ii}^*(\partial_{\mu}\Delta_{ii})V^{ii}_{\nu} \,, \label{eq:deltaconv}\\
    D^{\Delta}_{\mu\nu,geo} &=\text{Re}\sum_{\mk,i\neq j}2\frac{\abs{u_i}^2-\abs{v_i}^2}{E_i+E_j}\left[ u_j^* v_j(\Delta_{jj}\vecDot{\partial_{\mu}i}{j}-\Delta_{ii}\vecDot{\partial_{\mu}\bar{i}}{\bar{j}})V^{ji}_{\nu} 
        +u_j v_j^*(\Delta_{jj}^*\vecDot{\partial_{\mu}\bar{i}}{\bar{j}}-\Delta_{ii}^*\vecDot{\partial_{\mu}i}{j})\bar{V}^{ji}_{\nu} \right] \nonumber\\
        &=\text{Re}\sum_{\mk,i\neq j}\frac{\varepsilon_i}{E_iE_j(E_i+E_j)(\varepsilon_i-\varepsilon_j)}
        \left[ \abs{\Delta_{jj}}^2(V^{ij}_{\mu}V^{ji}_{\nu}+\bar{V}^{ij}_{\mu}\bar{V}^{ji}_{\nu})-\Delta_{ii}\Delta_{jj}^*\bar{V}^{ij}_{\mu}V^{ji}_{\nu}-\Delta_{ii}^*\Delta_{jj}V^{ij}_{\mu}\bar{V}^{ji}_{\nu} \right].\label{eq:deltageo}
    \end{align}
\end{widetext}
In the second equation of ~Eq.~\eqref{eq:deltaconv}, we have used the property that $V^{ii}=\bar{V}^{ii}$ and that $\vecDot{\partial_{\mu}i}{i}$ is purely imaginary. And in getting the second line of Eq.~\eqref{eq:deltageo}, we have employed the relation $\vecDot{\partial_{\mu}i}{j}=V^{ij}_\mu/(\varepsilon_i-\varepsilon_j)$ and $\vecDot{\partial_{\mu}\bar{i}}{\bar{j}}=\bar{V}^{ij}_\mu/(\bar{\varepsilon}_i-\bar{\varepsilon}_j)$. 

We first note that $D^{\Delta}_{\mu\nu,conv}$ appears even in single-band models, so long as the form factors of the intraband pairing function $\Delta_{ii}$ exhibits $\mk$-dependence, as was also noted in Ref.~\onlinecite{setty2023mechanism}. This term cannot be described by the physical processes sketched in Fig.~\ref{fig:sketch}. It measures the energy cost associated with the deformation of the Cooper pair wavefunction induced by supercurrent flow (or, real-space phase gradient). A gap function with $\mk$-dependence suggests that pairing takes place between electrons located on different sites. In such case, spatial phase variation can deform the Cooper pair wavefunction if the two electrons experience different phases. By contrast, no such deformation occurs for on-site Cooper pairing. 

We now focus on the geometric terms containing interband velocities. In close inspection, two types of processes can be identified. One is Cooper pair transfer from one band to another, as is associated with terms like $\Delta_{ii}^\ast\Delta_{jj}V^{ij}_\mu\bar{V}_\nu^{ji}$, the same as in Eq.~\eqref{eq:geo0} and Fig.~\ref{fig:sketch} (b). The other is associated with terms in the forms of 
\begin{equation}
    \abs{\Delta_{jj}}^2V^{ij}_\mu V^{ji}_\nu,~\abs{\Delta_{jj}}^2\bar{V}^{ij}_\mu \bar{V}^{ji}_\nu,~
    \ \text{and their conjugate.}
    \label{eq:processc}
\end{equation}
Note that the two elements of velocity operator appearing in each of these expressions share the same wavevector and are both electron (hole) type. These are unlike the Cooper pair tunneling terms in Eq.~\eqref{eq:geo0} where they are associated with opposite wavevectors, and where one is electron while the other is hole type. As a consequence, this geometric SW stems from virtual back-and-forth tunneling of a single electron between two different bands, as schematically depicted in Fig.~\ref{fig:sketch} (c). 

Importantly, this contribution remains finite even when only one of the bands develops superconductivity. Taking $\Delta_{11}\neq 0$ and $\Delta_{ii} =0$ for all $i\neq 1$, we arrive at
\begin{equation}
\begin{aligned}
D_{\mu\nu,geo}^{\Delta} =&\text{Re}\sum_{\mk,i\neq 1}\frac{\abs{\Delta_{11}}^2}{E_1(E_1+|\varepsilon_i|)}\frac{\varepsilon_i}{|\varepsilon_i|(\varepsilon_i-\varepsilon_1)}\times\\
&\left( V_{\mu}^{i1}V_{\nu}^{1i}+\bar{V}_{\mu}^{i1}\bar{V}_{\nu}^{1i} \right).
\end{aligned}
\label{eq:geoSingleband}
\end{equation}
Now consider a special case where band-1 is a flat band lying at the Fermi level, {\it i.e.}, $\varepsilon_1 \equiv 0$, while all other bands share the same dispersion $\varepsilon_i = \varepsilon_2$ for $i \neq 1$ and are well-separated from band-1 with $|\varepsilon_2| \gg |\Delta_{11}|$, one obtains,
\begin{equation}
\begin{aligned}
D_{\mu\nu,geo}^{\Delta} &\approx 2\sum_{\mk,i\neq 1} \abs{\Delta_{11}} \sqrt{n_{1}(1-n_{1})} \frac{\text{Re}\left(V_{\mu}^{i1}V_{\nu}^{1i}\right)}{\abs{\varepsilon_i}^2}\\
&\approx 2\sum_{\mk} |\Delta_{11}| \sqrt{n_{1}(1-n_{1})} \text{Re}\left(\mathcal{R}^1_{\mu\nu}\right)\,.
\end{aligned}
\label{eq:geoFlatband}
\end{equation}
In getting this we have applied the relation $\abs{\Delta_{11}}/E_1 = 2\sqrt{u_1^2v_1^2} = 2 \sqrt{n_{1}(1-n_{1})}$, where $n_{1} = |v_{1}|^2$ is the electron occupancy of band-1. Since $\text{Re}\left(\mathcal{R}^i_{\mu\nu}\right) > 0$, this SW is positive definite. This result agrees with previous studies that discuss superconductivity in flatbands~\cite{peottaSuperfluidityTopologicallyNontrivial2015,liangBandGeometryBerry2017,xieTopologyBoundedSuperfluidWeight2020}. As the conventional SW vanishes in flatbands, this geometric SW Eq.~\eqref{eq:geoFlatband} provides the sole mechanism to maintain phase coherence in such superconductors. A pictorial interpretation goes as follows: the paired immobile electrons in the flat band can still `move' via virtual interband tunneling, thereby enabling Cooper pair flow.


Finally, if we consider real pairing order parameters, Eq.~\eqref{eq:deltageo} can be simplified to
\begin{equation}
D_{\mu\nu,geo}^{\Delta} =\sum_{\mk ,i\neq j}\frac{\varepsilon_i\Delta_{jj}(\Delta_{jj}-\Delta_{ii})}{E_iE_j(E_i+E_j)(\varepsilon_i-\varepsilon_j)}(V_{\mu}^{ij}V_{\nu}^{ji}+V_{\mu}^{ji}V_{\nu}^{ij}).
\label{eq:geo1trs}
\end{equation}
This term can take either positive or negative values, depending on the microscopic details in the band dispersion and the order parameter configuration on the bands. It vanishes in the special case of a two-band model with $\Delta_{11}=\Delta_{22}$. 

Combining Eqs.~\eqref{eq:geo0trs} and \eqref{eq:geo1trs} gives the total geometric SW in the absence of interband pairing. 
\begin{equation}
\begin{aligned}
D_{\mu\nu,geo} =& \sum_{\mk,i<j}\frac{1}{E_iE_j(E_i+E_j)}V_{\mu}^{ij}V_{\nu}^{ji}\times\\
&\left[ 4\Delta_{ii}\Delta_{jj}+2\frac{(\varepsilon_i\Delta_{jj}+\varepsilon_j\Delta_{ii})(\Delta_{jj}-\Delta_{ii})}{\varepsilon_i-\varepsilon_j}\right].
\end{aligned}
\label{eq:totalgeo}
\end{equation}
Generally, a finite total geometric SW is generated between any pair of intraband order parameters $\Delta_{ii}$ and $\Delta_{jj}$. Nonetheless, it can be checked that for a two-band system, $D_{\mu\nu,geo}$ vanishes if $\Delta_{11}=-\Delta_{22}$.

\section{Interband-pairing induced superfluid weight}\label{sec:inter}
The derivation in the previous section has ignored interband pairing. In reality, there could be cases where multiple bands lie in proximity and where the interaction is strong compared to the bandwidth and band separation, such as in flatbands. In this case, interband pairing may naturally emerge in strength and is thus no longer negligible. In the following, we turn to the SW associated with interband pairing, illustrating the essential physics through a two-band model. \par

We consider interband spin-singlet pairing between electrons at opposite wavevectors, {\it i.e.}, $\Delta_{ij} c_{i\bar{\mk} \downarrow}c_{j \mk \uparrow}$. Let us first focus on a simpler scenario with only interband pairing in the pairing matrix: $\Delta_{11}=\Delta_{22}=0$ and $\Delta_{12}=\Delta_{21}$. The two-band BdG Hamiltonian reads,
\begin{equation}
    \mathcal{H}^{\bf{BdG}}_{\mk}=
    \begin{pmatrix}
        \varepsilon_1 & 0 & 0 & \Delta_{12}\\
        0 & \varepsilon_2 & \Delta_{21} & 0\\
        0 & \Delta_{21}^* & -\varepsilon_1 & 0\\
        \Delta_{12}^* & 0 & 0 & -\varepsilon_2
    \end{pmatrix}.
\end{equation}
In this case, the BdG Hamiltonian can be decomposed into two decoupled $2\times2$ subblocks. The Bogoliubov spectra of one of the subblocks follows as
\begin{equation}
\begin{aligned}
        E^{\pm}_{\alpha}&=\frac{\varepsilon_1-\varepsilon_2}{2}\pm g(\Delta_{12})\,, 
\end{aligned}
\end{equation}
where $g(x)=\sqrt{\frac{1}{4}(\varepsilon_1+\varepsilon_2)^{2}+\abs{x}^2}$. The corresponding quasiparticle wavefunctions are given respectively by  $u_\alpha\rvec{+,1}+v_\alpha\rvec{-,\bar{2}}$ and $-v_\alpha^\ast\rvec{+,1}+u_\alpha^\ast\rvec{-,\bar{2}}$, where the coefficients satisfy
\begin{equation}\label{eq:uvinter}
    \begin{aligned}
        u_\alpha v_\alpha^*&=\frac{\Delta_{12}}{2g(\Delta_{12})} \,.
    \end{aligned}
\end{equation}
Similarly, the other subblock has
\begin{equation}
\begin{aligned}
        E^{\pm}_{\beta}&=\frac{\varepsilon_2-\varepsilon_1}{2}\pm g(\Delta_{21})\,, 
\end{aligned}
\end{equation}
with wavefunctions $u_\beta \rvec{+,2}+v_\beta\rvec{-,\bar{1}}$ and $-v_\beta^\ast\rvec{+,2}+u_\beta^\ast\rvec{-,\bar{1}}$, and the coefficients satisfying
\begin{equation}\label{eq:uvinter1}
    \begin{aligned}
        u_\beta v_\beta^*&=\frac{\Delta_{21}}{2g(\Delta_{21})} \,.
    \end{aligned}
\end{equation}
Because $\Delta_{12}=\Delta_{21}$, one in fact has $u_\alpha=u_\beta$ and $v_\alpha=v_\beta$. However, we distinguish the two sets of coefficients to allow for a clear distinction of physical processes involving different subsets of Hilbert space in subsequent analyses. 

The SW Eqs.~\eqref{eq:sw_Dkin} and \eqref{eq:sw_Ddelta} can then be obtained by direct substitution of the above quantities. As in the intraband-pairing-only scenario, the kinetic-gradient contribution of Eq.~\eqref{eq:sw_Dkin} can be decomposed to conventional and geometric parts. The conventional part follows as
\begin{equation}\label{eq:InterbandSWConv}
    \begin{aligned}
            D^{\text{kin}}_{\mu\nu,conv}&=2\sum_{\mk}C_{1122}^{\ti{++--}}\left[ V^{11}_{\mu}\bar{V}^{22}_{\nu}+(\mu\!\leftrightarrow\! \nu) \right]\\
            &+2\sum_{\mk}C_{2211}^{\ti{++--}}\left[V^{22}_{\mu}\bar{V}^{11}_{\nu}+(\mu\!\leftrightarrow\! \nu) \right] \,,
    \end{aligned}
\end{equation}
with
\begin{equation}\label{eq:InterbandSWCoeff}
    \begin{aligned}
                C_{1122}^{\ti{++--}}&=-\frac{f(E_{\alpha}^{\ti{+}})\!-\!f(E_{\alpha}^{\ti{-}})}{g(\Delta_{12})}\abs{u_{\alpha}v_{\alpha}}^2,\\
        C_{2211}^{\ti{++--}}&=-\frac{f(E_{\beta}^{\ti{+}})\!-\!f(E_{\beta}^{\ti{-}})}{g(\Delta_{21})}\abs{u_{\beta}v_{\beta}}^2.
    \end{aligned}
\end{equation}
Utilizing Eqs.~\eqref{eq:uvinter} and \eqref{eq:InterbandSWCoeff}, we arrive at a $D^{\text{kin}}_{\mu\nu,conv}$ containing the following expressions in the multiple terms of the integrand of Eq.~\eqref{eq:InterbandSWConv}:
\begin{equation}
    \abs{\Delta_{12}}^2V^{11}_{\mu}\bar{V}^{22}_{\nu},
    \ \abs{\Delta_{21}}^2V^{22}_{\mu}\bar{V}^{11}_{\nu},
    \ \text{and}\ \mu\leftrightarrow \nu.
    \label{eq:processd}
\end{equation}
In analogy to the interpretation in the previous section, Eq.~\eqref{eq:InterbandSWConv} can be thought of as originating from the transfer of an interband Cooper pair to itself by virtue of the intraband velocities, as sketched in Fig.~\ref{fig:sketch}~(d). Since $E^+_{\alpha(\beta)} \geq E^-_{\alpha(\beta)}$, both $C_{1122}^{\ti{++--}}$ and $C_{2211}^{\ti{++--}}$ are non-negative. This leads to an interesting observation that the sign of this SW depends on the relative sign of $V^{11}$ and $V^{22}$. A negative $D^{\text{kin}}_{\mu\nu,conv}$ results for a model with a pair of particle-like and hole-like bands. In the absence of the geometric contribution ({\it i.e.}, $V^{12}\equiv 0$), a negative $D^{\text{kin}}_{\mu\nu,conv}$ implies that such a multiband model with only uniform interband pairing is inherently unstable, and that a PDW state may thus be likely to form. \par

The geometric contribution of Eq.~\eqref{eq:sw_Dkin} is given by
\begin{equation}
    \begin{aligned}
            D^{\text{kin}}_{\mu\nu,geo}&=2\sum_{\mk}C_{1212}^{\ti{++--}}\left[V^{12}_{\mu}\bar{V}^{12}_{\nu}+(\mu\leftrightarrow \nu)\right] \\
            &+2\sum_{\mk}(C_{1212}^{\ti{++--}})^*\left[V^{21}_{\mu}\bar{V}^{21}_{\nu}+(\mu\leftrightarrow \nu)\right],\\
        \label{eq:Dinterkingeo}
    \end{aligned}
\end{equation}
with
\begin{equation}
    \begin{aligned}
        C_{1212}^{\ti{++--}}&=\sum_{\sigma=\pm}\frac{f(E_{\alpha}^{\sigma})\!-\!f(E_{\beta}^{\sigma})}{E_{\alpha}^{\sigma}\!-\!E_{\beta}^{\sigma}}u_{\alpha}^*v_{\alpha}u_{\beta}v_{\beta}^* .
    \end{aligned}
\end{equation}
Taken together, we obtain a $D^{\text{kin}}_{\mu\nu,geo}$ containing the following expressions in the integrand
\begin{equation}
    \Delta_{12}^*\Delta_{21}V^{12}_{\mu}\bar{V}^{12}_{\nu},
    \ \Delta_{21}^* \Delta_{12}V^{21}_{\mu}\bar{V}^{21}_{\nu},
    \ \text{and}\ \mu\leftrightarrow \nu.
    \label{eq:processe}
\end{equation}
These terms thus represent Cooper pair transfer between two interband pairs $\Delta_{12}$ and $\Delta_{21}$, which is implemented through interband velocity as shown in Fig.~\ref{fig:sketch} (e).\par

The $\Delta$-gradient contribution Eq.~\eqref{eq:sw_Ddelta} consists of eight terms associated with the following $\mk$-dependent coefficients:
\begin{equation}
    \begin{aligned}
        &C^{\ti{+-++}}_{1121},~C^{\ti{+-++}}_{1211},~C^{\ti{+-++}}_{2122},~C^{\ti{+-++}}_{2212}\\
        &C^{\ti{+---}}_{1222},~C^{\ti{+---}}_{1112},~C^{\ti{+---}}_{2111},~C^{\ti{+---}}_{2221}\ .
    \end{aligned}
\end{equation}
We focus on the $C^{\ti{+-++}}_{1121}$ and $C^{\ti{+-++}}_{1211}$ terms since they are representative of all eight terms. The $C^{\ti{+-++}}_{1121}$ term involves the integral of the following
\begin{equation}
    \begin{aligned}
    -2\text{Re}\left\{C_{1121}^{\ti{+-++}}\left( 
\Delta_{12}\vecDot{\partial_{\mu}\bar{2}}{\bar{1}}-\Delta_{21}\vecDot{\partial_{\mu}1}{2} \right)V^{21}_{\nu} \right\},
    \end{aligned}
\end{equation}
where
\begin{equation}
    \begin{aligned}
        C_{1121}^{\ti{+-++}}&=\sum_{\sigma=\pm}\sigma\frac{f(E_{\alpha}^{+})-f(E_{\beta}^{\sigma})}{E_{\alpha}^{+}-E_{\beta}^{\sigma}}\abs{u_{\alpha}}^2u_{\beta}^*v_{\beta}\\
        &+\sum_{\sigma=\pm}\sigma\frac{f(E_{\alpha}^{-})-f(E_{\beta}^{\sigma})}{E_{\alpha}^{-}-E_{\beta}^{\sigma}}\abs{v_{\alpha}}^2u_{\beta}^*v_{\beta}.
    \end{aligned}
\end{equation}
Using Eqs.~\eqref{eq:Velocityij} and \eqref{eq:uvinter} we identify two types of terms. The first one involves $\Delta_{21}^*\Delta_{12}\bar{V}^{21}_{\mu}V^{21}_{\mu}$, which depicts the same interband pair hopping as in Eq.~\eqref{eq:Dinterkingeo}. The second one involves
\begin{equation}
    \abs{\Delta_{21}}^2V^{12}_{\mu}V^{21}_{\nu}\,.
    \label{eq:processf}
\end{equation}
Similar to Eq.~\eqref{eq:processc}, this contribution is described by a back-and-forth one-electron interband tunneling as sketched in Fig.~\ref{fig:sketch} (f).  \par
In addition, coefficient $C^{\ti{+-++}}_{1211}$ correspond to the following integral
\begin{equation}\label{eq:partialD12}
    -2\text{Re}\left[ C^{\ti{+-++}}_{1211}(\partial_{\mu}\Delta_{12})V^{11}_{\nu} \right],
\end{equation}
where
\begin{equation}
    \begin{aligned}
        C^{\ti{+-++}}_{1211}&=\frac{f(E_{\alpha}^+)-f(E_{\alpha}^-)}{2g(\Delta_{12})}\left( \abs{v_{\alpha}}^2-\abs{u_{\alpha}}^2 \right)u_{\alpha}^*v_{\alpha}\\
        &+\frac{f(E_{\beta}^+)-f(E_{\beta}^-)}{2g(\Delta_{21})}\left( \abs{v_{\beta}}^2-\abs{u_{\beta}}^2 \right)u_{\beta}^*v_{\beta}.
    \end{aligned}
\end{equation}
Again using Eq.~\eqref{eq:uvinter}, one obtains expressions containing  $(\varepsilon_{1}+\varepsilon_{2})\Delta_{12}^*\partial_{\mu}\Delta_{12}V^{11}_{\nu}$. They are of similar origin to the geometric SW given by Eq.~\eqref{eq:deltaconv} and are also not plotted in Fig.~\ref{fig:sketch}.\par
Finally, we also consider two-band models containing both intraband and interband pairings. Treating interband pairing as a perturbation, we perform a first-order perturbative expansion of the quasiparticle wavefunctions. Recall that the unperturbed solutions are given by
$\pm E_i$ and $\rvec{\psi_{i}^{\pm}}~(i=1,2)$. The first order correction to the energy vanishes, while the wavefunction expands as follows:
\begin{equation}\label{eq:1thpsi}
    \begin{aligned}
        \rvec{\psi_{1}^{+}}'&=\frac{1}{E_1-E_2}(u_2^*v_1\Delta_{21}+v_2^*u_1\Delta_{12}^*)\rvec{\psi_2^+}\\
        &+\frac{1}{E_1+E_2}(u_2u_1\Delta_{12}^*-v_2v_1\Delta_{21})\rvec{\psi_2^-},\\
        \rvec{\psi_{1}^{-}}'&=\frac{1}{E_1+E_2}(v_2^*v_1^*\Delta_{12}^*-u_2^*u_1^*\Delta_{21})\rvec{\psi_2^+}\\
        &+\frac{1}{E_1-E_2}(v_2u_1^*\Delta_{21}+u_2v_1^*\Delta_{12}^*)\rvec{\psi_2^-}.
    \end{aligned}
\end{equation}
The full SW can then be obtained by substituting Eq.~\eqref{eq:1thpsi} into Eqs.~\eqref{eq:sw_Dkin} and \eqref{eq:sw_Ddelta}. In addition to terms already discussed in the above analyses, we obtain new contributions of linear order in $\Delta_{12}~(\Delta_{21})$ deriving from Eq.~\eqref{eq:sw_Dkin}. Their integrands contain expressions of the following form 
\begin{equation}
    \begin{aligned}
        &\frac{1}{E_1(E_1-E_2)}\abs{v_1}^2\abs{u_2}^2u_1v_1^*\Delta_{21}^*V^{21}_{\mu}\bar{V}^{11}_{\nu}\\
        =&\frac{1}{2E_1^2(E_1-E_2)}\abs{v_1}^2\abs{u_2}^2\Delta_{21}^\ast \Delta_{11}V^{21}_{\mu}\bar{V}^{11}_{\nu}.
        \label{eq:processg}
    \end{aligned}
\end{equation}
They describe the transfer of an intraband Cooper pair to an interband Cooper pair or vice versa, facilitated by a joint action of intraband and interband velocities as shown in Fig.~\ref{fig:sketch}~(g). Here, the product of the velocity elements, $V^{21}_{\mu}V^{11}_{\nu}$, cannot be turned into the gauge-invariant geometric tensor even in the flatband limit. Nonetheless, gauge invariance is ensured by the multiplication of interband velocity by interband pairing in this expression. 

\section{Model calculation}\label{sec:model}
Having derived the SW for general multiband superconductors, we present in this section a numerical calculation of the SW for a two-band model consisting of an $s$- and a $d_{xy}$-orbital on each site of a square lattice. This simple topologically trivial model was initially studied in Ref.~\onlinecite{chenPairDensityWave2023}. Here, we shall explicitly separate the geometric contribution to the SW and study how it varies with the multiband pairing configuration. 

\begin{figure}[b]
    \includegraphics[width=\columnwidth]{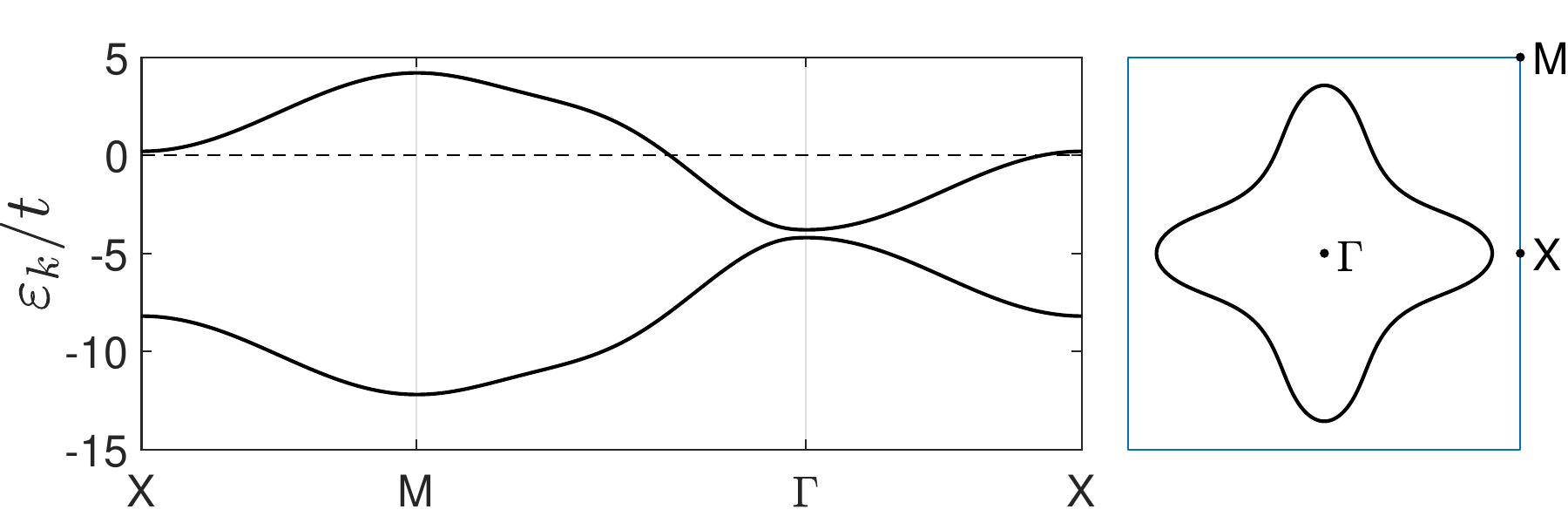}
    \caption{The band structure and Fermi surface of the $s$-$d_{xy}$ two-band model with parameters $(t_s,t_d,t',\mu_s,\mu_d)=(1,-1,1,-0.2,8.2)t$.}
    \label{fig:sdxy_band}
\end{figure}

The kinetic Hamiltonian of the model written in the basis $(c_{s\mk},c_{d\mk})^T$ follows as
\begin{equation}
\tilde{\mathcal{H}}_{\mk}=
\begin{pmatrix}
    \xi_{s\mk} & \lambda_{\mk}\\
    \lambda_{\mk}^{*} & \xi_{d\mk}
\end{pmatrix},
\end{equation}
where $\xi_{a\mk}=-2 t_a\left(\cos k_x+\cos k_y\right)-\mu_a,\ (a=s, d)$ and $\lambda_{\mk}=4t'\sin{k_{x}}\sin{k_{y}}$. Figure \ref{fig:sdxy_band} shows the band structure and Fermi surface of the model for a set of tight-binding parameters used in the following calculations. \par 

In this section, we perform calculations in the band-basis representation. The total SW can be obtained from the band-basis version of Eq.~\ref{eq:Doriginal}, while the geometric contribution to the SW is distinguished by whether it involves the interband velocity (or interband Berry connection) as our preceding analyses have outlined. Since the geometric SW becomes significant only at strong pairing, we shall push to rather large $\Delta$ in our numerics. Admittedly, the validity of the mean-field BdG formalism is questionable in this limit. However, it is reasonable to expect such calculations to be able to capture at least some qualitative features in strong coupling.   

\begin{figure}
	\hspace{-0.1\columnwidth}
	\includegraphics[width=0.8\columnwidth]{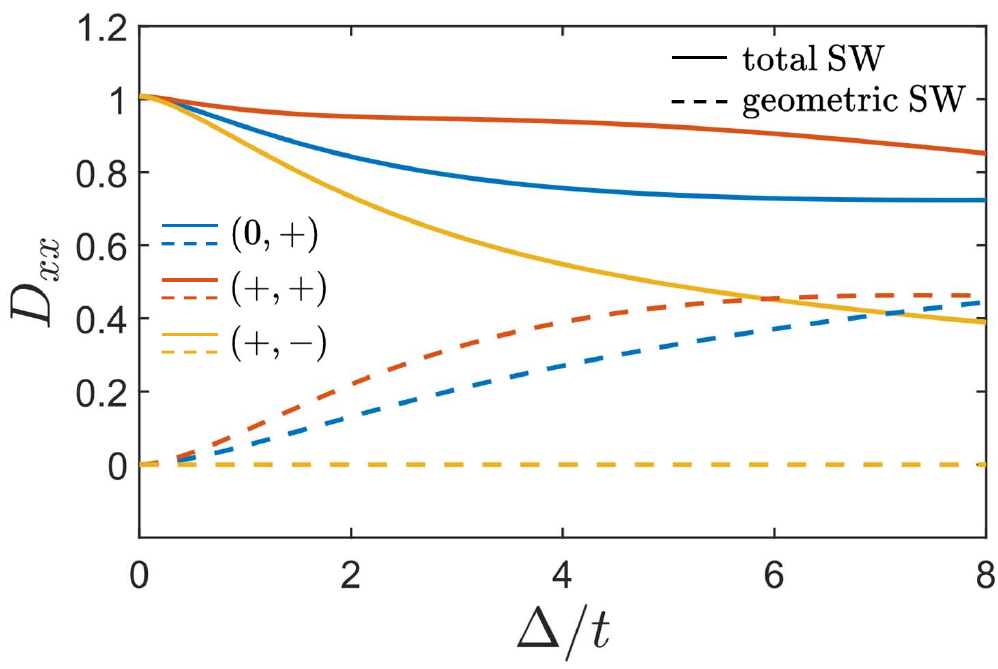}
	\caption{The SW of the $s$-$d_{xy}$ model given in Fig.~\ref{fig:sdxy_band} in three different configurations of intraband pairing in Eq.~\eqref{eq:intraconfig}. The solid and dashed curves depict the total SW and the geometric contribution, respectively.}
 \label{fig:sdxy-intra}
\end{figure}

We first consider the following spin-singlet superconducting states without interband pairing:
\begin{equation}
\begin{aligned}
\bDelta^{(0,+)}&=
\begin{pmatrix}
    0 & 0\\
    0 & \Delta
\end{pmatrix},\\
\bDelta^{(+,+)}&=
\begin{pmatrix}
    \Delta & 0\\
    0 & \Delta
\end{pmatrix},\\
\bDelta^{(+,-)}&=
\begin{pmatrix}
    \Delta & 0\\
    0 & -\Delta
\end{pmatrix}.\\
\end{aligned}
\label{eq:intraconfig}
\end{equation}
Figure \ref{fig:sdxy-intra} plots the total SW $D_{xx}$ and the corresponding geometric contribution as a function of the pairing amplitude. Consistent with the discussions following Eq.~\eqref{eq:geo0}, in all cases the geometric SW diminishes approaching the weak-pairing limit. Generally, the geometric SW is sensitive to the pairing configuration on the bands. For the $(0,+)$ and $(+,+)$ configurations, the geometric contribution first increases monotonically with increasing pairing strength, before eventually saturating as the gap amplitude reaches the bandwidth. The geometric SW vanishes in the $(+,-)$ configuration, as was predicted in the end of Sec.~\ref{sec:SWdelta}. 

\begin{figure}
	\hspace{-0.1\columnwidth}
	\includegraphics[width=0.8\columnwidth]{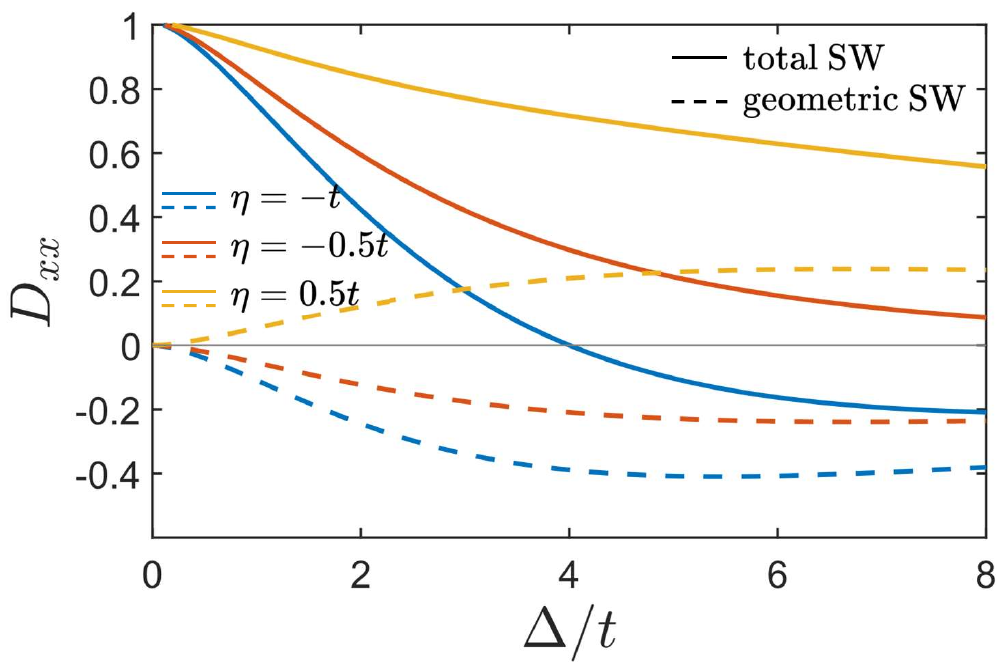}
	\caption{The SW of the $s$-$d_{xy}$ model given in Fig.~\ref{fig:sdxy_band} for superconducting states with opposite intraband pairing and also involving interband pairing, as shown in Eq.~\eqref{eq:InterbandPairing}. The solid and dashed curves depict the total SW and the geometric contribution, respectively.}
 \label{fig:sdxy-inter}
\end{figure}

A natural next step is to investigate the effect of interband pairing. We shall focus on the more interesting scenario with opposite intraband pairings. The pairing matrix reads, 
\begin{equation}\label{eq:InterbandPairing}
\bDelta=
\begin{pmatrix}
    1 & \eta\sin{k_x}\sin{k_y}\\
    \eta \sin{k_x}\sin{k_y} & -1
\end{pmatrix}\Delta.
\end{equation}
Here, $\eta$ is a control parameter used to tune the strength and phase of the interband pairing. The symmetry of the interband pairing is determined in tandem with the symmetry of the two Bloch bands. In the present model, an interband pairing having the basis function of $\Delta_{12,\mk} \propto k_xk_y$ can be checked to remain intact under all symmetry operations of the underlying lattice, and is hence of the same symmetry channel as the intraband $s$-wave pairing. In the calculations, we take $\Delta_{12,\mk} \propto \sin k_x\sin k_y$, without loss of generality. 

Figure~\ref{fig:sdxy-inter} shows the numerical results for a representative set of $\eta$. Intriguingly, the geometric SW is exactly opposite for $\eta=0.5t$ and $\eta=-0.5t$. As we saw above as well as in Eq.~\eqref{eq:totalgeo}, the geometric contribution primarily associated with intraband pairings vanishes if $\Delta_{11}=-\Delta_{22}$. Hence, the finite opposite geometric SW here must stem from Eq.~\eqref{eq:processg}, which changes sign when $\Delta_{12}$ does. Overall, increasingly stronger interband pairing gives rise to increasingly larger magnitude of geometric SW. Meanwhile, with a pair of particle- and hole-like bands in the present model (Fig.~\ref{fig:sdxy_band}), the conventional contribution associated with Eq.~\eqref{eq:processd} (Fig.~\ref{fig:sketch} (d)) also becomes more negative with increasing $|\Delta_{12}|$. Put together, a striking negative total SW is achieved at strong intraband pairing as well as sufficiently negative $\eta$, $\eta=-t$. A negative total SW, or negative phase stiffness, is unphysical, suggesting that such a superconducting state is unstable. On the other hand, it implies a tendency to simultaneously develop spatial phase modulation. This provides a natural intrinsic route to form PDW, as has been demonstrated in several preliminary studies~\cite{chenPairDensityWave2023,wang2024quantum,jiangPairDensityWaves2023}. 

\section{Summary}\label{sec:summary}
In this study, we have provided a microscopic interpretation of the origin of the quantum-geometry-induced SW in multiband superconductors. We analytically derived the SW of multiband models with only intraband Cooper pairing as well as those also involving interband pairing, and elucidated the microscopic processes underlying each of the contributing terms. The geometric SW was found to relate closely to tunneling processes enabled by interband velocity (or interband Berry connection). Two types of such processes are identified. One is akin to the Cooper pair transfer between the bands, which influences the SW by inducing effective Josephson coupling between different pairing order parameters. The other is single-electron back-and-forth tunneling between different bands. For a superconducting flatband well-separated from other bands, this latter process stimulates effective transport of the purported immobile electrons that form Cooper pairs.

We further calculated the SW of an exemplary two-band model, explicitly demonstrating how quantum geometry may affect the nature of the superfluidity, in a manner that is sensitive to the configuration of pairing on the multiple bands. The geometric effect can either enhance or reduce the total SW. Highlighted prominently is a case with negative total SW, which benefits in part from a negative geometric contribution. A strongly suppressed or negative total SW may constitute a novel intrinsic mechanism for the formation of PDW states~\cite{chenPairDensityWave2023,wang2024quantum,jiangPairDensityWaves2023}. 

Finally, as the geometric SW is more pronounced in the strong pairing limit, our results are more relevant to narrow-band and flatband superconductors, or conversely, in dispersive systems with very strong interactions. Looking forward, it is worthwhile to investigate the quantum geometric effects using more sophisticated many-body approaches, beyond the mean-field formalism employed in this work. Further, as another interesting future direction, the effective interband Josephson coupling induced by quantum geometry may play a role in the collective phase fluctuations, such as Goldstone and Leggett modes~\cite{leggett1966number}, of multiband superconductors or superfluids.

\begin{acknowledgments}
We acknowledge helpful discussions with Wei-Qiang Chen, Zhiqiang Wang and Chang-Ming Yue. This work is supported by NSFC under Grants No.~12374042 and No.~11904155, the Guangdong Science and Technology Department under Grant 2022A1515011948, and a Shenzhen Science and Technology Program (Grant No.~KQTD20200820113010023). 
\end{acknowledgments}


\bibliography{QG_analytic}

\end{document}